# Exploring the Cybersecurity-Resilience Gap: An Analysis of Student Attitudes and Behaviors in Higher Education

Steve Goliath[0009-0007-9335-9184], Pitso Tsibolane[0000-0002-7888-1181] and Dirk Snyman[0000-0001-7360-3214]

Department of Information Systems
University of Cape Town, Rondebosch, South Africa
pitso.tsibolane@uct.ac.za

**Abstract.** Cyberattacks frequently target higher educational institutions, making cybersecurity awareness and resilience critical for students. However, limited research exists on cybersecurity awareness, attitudes, and resilience among students in higher education. This study addresses this gap using the Theory of Planned Behavior as a theoretical framework. A modified Human Aspects of Information Security Questionnaire was employed to gather 266 valid responses from undergraduate and postgraduate students at a South African higher education institution. Key dimensions of cybersecurity awareness and behavior, including password management, email usage, social media practices, and mobile device security, were assessed. A significant disparity in cybersecurity awareness and practices, with postgraduate students demonstrating superior performance across several dimensions was noted. This research postulates the existence of a Cybersecurity-Education Inflection Point during the transition to postgraduate studies, coined as the Cybersecurity-Resilience Gap. These concepts provide a foundation for developing targeted cybersecurity education initiatives in higher education, particularly highlighting the need for earlier intervention at the undergraduate level.

## 1      Introduction

The COVID-19 pandemic has accelerated the adoption of Information and Communication Technology (ICT) in educational institutions, leading to a significant shift towards online learning. This digital transformation has made higher education institutions attractive targets for cyberattacks [1]. The expanding cyberworld necessitates the widespread adoption of cybersecurity measures, as students, faculty, and staff remain primary targets [2].

The cost of data breaches and cyberattacks can be devastating, with the global average cost reaching 4.88 million USD in 2024 [3]. Email remains the most common vector for malware attacks, with 94% of organizations reporting email-based cyber-attacks [4].

Educational institutions face unique challenges in cybersecurity due to their valuable data assets (such as financial and personal information and research data) and limited IT infrastructure to defend against attacks. Malware and phishing attacks are the most common threats, with educational institutions ranking fifth globally in cybercrime incidents [5]. Ransomware has emerged as a devastating threat, with over 1,000 academic institutions victimized since 2019 [6]. Yet, there is a notable lack of research on cybersecurity awareness among undergraduate and postgraduate students in South African higher education institutions. Therefore, this study aims to address this research gap by evaluating attitudes toward cybersecurity awareness behaviors among students at a South African institution of higher education. Using the Human Aspects of Information Security Questionnaire (HAIS-Q) [7], the following research question was investigated:

*What are the attitudes of higher education students at various levels of study towards cybersecurity awareness?*

The remainder of this paper is structured as follows: Section 2 presents a comprehensive literature review on cybersecurity awareness in higher education. Section 3 describes the methodology employed in this study, including using the HAIS-Q instrument. Section 4 presents the results of our analysis, followed by a discussion of these findings in Section 5. Finally, Section 6 concludes the paper with implications for practice and suggestions for future research.

## 2 Literature review

This section reviews existing literature on cybersecurity awareness and attitudes in various contexts. It explores critical dimensions of cybersecurity practices, including password management, email and internet usage, social media behavior, mobile device security, and incident reporting. Additionally, the review examines studies on human attitudes toward security awareness behaviors, providing a theoretical foundation for understanding the psychological factors influencing cybersecurity practices.

### 2.1 Cybersecurity awareness in South Africa

Studies on cybersecurity awareness in South Africa reveal concerning trends. Ross and Rasool [8] found students feeling unsafe on campus, while Chandarman and van Niekerk [9] identified high risks of cyberattacks among college students due to misaligned awareness variables. There is a lack of cybersecurity awareness and collaboration among policymakers and an absence of cybercrime-related policing and awareness programs by the South African Police Service [16].

### 2.2 Study dimensions

This study focuses on critical dimensions of cybersecurity, including *password management*, *email* and *internet usage*, *social media behavior*, *mobile device security*, and *incident reporting* [7]. It also examines security awareness behaviors and attitudes, building on existing research in cybersecurity, cyberbullying, and cyberspace.

*Password management and Phishing*—Recent studies on university students' cybersecurity awareness reveal common vulnerabilities. Nigerian students need education in password management and email phishing [10]. Another study identified password management and public computer use as key risk areas among Hungarian students [11]. Sudanese undergraduate students had minimal awareness of password protection and antivirus use and lacked an understanding of various cybersecurity threats [1].

*Email usage*: Recent studies highlight students' vulnerability to phishing attacks. Burita, Klaban, and Racil [12] developed a system to train students and faculty about phishing threats based on an analysis of evolving phishing email tactics. Seventy percent (70%) of students at a Nigerian university were susceptible to cyberattacks due to poor cybersecurity awareness, particularly regarding public computer use and phishing recognition [13]. Sixty percent (60%) of tertiary education students clicked on links in simulated phishing emails [14].

*Internet usage* – Several studies highlight cybersecurity awareness gaps among university students [15]. Undergraduate students in Saudi Arabia, lacked basic cybersecurity knowledge, recommending active measures to improve awareness of password management, email security, and social media usage [16]. Similarly, a study of 1,231 students at a South African private tertiary institution identified misalignments between cybersecurity knowledge, self-perception of skills, and actual behavior, indicating high vulnerability to cyberattacks [9].

*Social media use* – A mixed-methods study of 100 online students revealed varying perceptions of cyberbullying and hacking on platforms like WhatsApp and Instagram [17]. Despite active academic use, social media remains vulnerable to credential theft. Advanced hacking techniques underscore the importance of robust password management and avoiding sharing credentials with unknown users. Previous research categorized cyberattack motivations as emotions, financial gains, entertainment, hacktivism, espionage, cyberwarfare, and job proficiency [18].

*Mobile device use* – Studies highlight the limitations of traditional security measures against Trojan and virus threats targeting smartphones [19]. Venter et al. [20] also emphasize the growing need for smartphone cybersecurity education in South Africa due to rising cyberattacks, smartphone adoption, and usage among children.

*Incident reporting* – A qualitative study of 15 teachers from five private schools in South Africa revealed a need for comprehensive cybersecurity awareness and program implementation to protect students from cyberbullying, online fraud, inappropriate content, and internet addiction [21]. Another study identified over 2,300 phishing emails targeting students and staff between 2010 and 2023 at Cornell University [22]. One in three individuals fell victim, often by clicking on malicious links. The study revealed authority and scarcity as common social engineering tactics attackers use.

## 2.3 Security awareness behavior and attitude

Human behavior is critical in cybersecurity, often leading to successful attacks. Despite its importance, cybersecurity education often overlooks the influence of attitudes on behavior [23]. Younger generations exhibit more secure digital practices than older ones, highlighting the impact of age and engagement [31].

Previous research on Hungarian and Vietnamese university students revealed a lack of in-depth knowledge of information security among many respondents [24]. This study explores the relationship between attitudes and cybersecurity behaviors among university students. The Theory of Planned Behavior (TPB) provides a theoretical framework for understanding how attitudes influence intentions and behavior. While TPB focuses on attitudes, subjective norms, and perceived behavioral control, this study will explore the specific impact of attitudes on cybersecurity behaviors. This research was undertaken as part of the research component of an honors degree at a South African university and was therefore limited in scope. Subsequent related work will see the other constructs included.

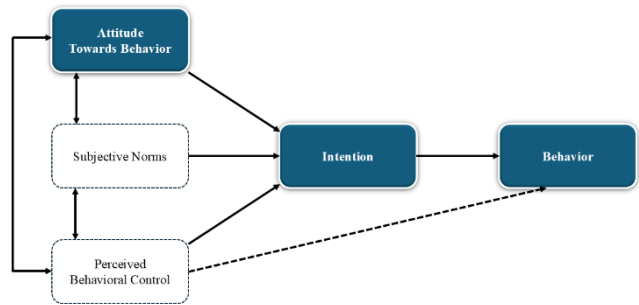

**Fig. 1.** Theory of planned behavior [25]

## 3　Methodology

The study adopted a positivist philosophy with a deductive approach, aligning with quantitative research methods. Following the TPB, hypotheses were developed and tested to validate or reject them based on collected data.

A cross-sectional study design was used, spanning eight months in 2023, with all institutional ethical approval. The target population was students, where probability sampling was applied, and a sample size of 380 was calculated. Students aged 18 and above were eligible to participate, with their age and study level recorded.

Data was collected using a modified version of the Human Aspects of Information Security Questionnaire (HAIS-Q), focusing on six key areas: *password management*, *email usage*, *internet usage*, *social media use*, *mobile device use*, and *incident reporting.* Closed-ended questions on a five-point Likert scale were used (*1. Strongly disagree* to *5. Strongly agree*). Questions were posed from both a positive and negative security perspective. Negatively phrased questions were assessed on a reversed scale, i.e., *5. Strongly disagree* down to *1. Strongly agree*. Ethical approval was obtained, and informed consent was collected. The questionnaire was distributed online via university communication channels, with respondents given the freedom to exit at any point. The survey took approximately ten minutes to complete, with all data treated confidentially.

Data cleaning was performed in Microsoft Excel, followed by analysis using Statistica (v. 10). Of the 285 responses received, seven were incomplete, and those from respondents over 40 were excluded, resulting in a final sample size of $N=266$. The data was organized with coded and labeled variables, and descriptive and inferential statistical techniques were employed. Hypotheses testing was conducted as follows:

**Null Hypothesis (H0):** No statistically significant difference exists in cybersecurity attitudes across students categorized by study level (undergraduate vs. postgraduate) or age. **Alternative Hypothesis (H1):** A statistically significant difference exists in cybersecurity attitudes across students categorized by study level or age.

## 4 Results

### 4.1 Descriptive statistics

Table 1 summarizes the demographic information. The highest percentage of participants were between the ages of 20-25 years old (~42%), while 158 (~60%) were undergraduates and 108 (~40%).

**Table 1.** Descriptive statistics about the respondents

| Demographic | Metric | Frequency | Percentage |
|---|---|---|---|
| Age | 15 - 20 years old | 87 | 32.71 % |
| | 21 - 25 years old | 112 | 42.11 % |
| | 26 - 30 years old | 35 | 13.16 % |
| | 31 - 35 years old | 18 | 6.77 % |
| | 36 - 40 years old | 14 | 5.26 % |
| Study Level | Undergraduate (UG) | 158 | 59.40% |
| | Postgraduate (PG) | 108 | 40.60% |

### 4.2 Focus Area descriptive statistics

Table 2 outlines the means and standard deviation for the responses received for each question for each focus area.

**Table 2.** Descriptive statistics according to the focus areas

| Focus area | Question label (^ indicates a negatively worded question) | Mean | Std Dev |
|---|---|---|---|
| Password management | PM1^ | 4.07 | 1.04 |
| | PM2 | 4.08 | 1.08 |
| | PM3^ | 4.27 | 0.84 |
| Email usage | EU1^ | 3.70 | 1.08 |
| | EU2^ | 4.83 | 0.41 |
| | EU3 | 3.89 | 1.02 |

| Focus area | Question label (^ indicates a negatively worded question) | Mean | Std Dev |
|---|---|---|---|
| Internet usage | IU1 | 3.91 | 0.92 |
|  | IU2 | 4.33 | 0.63 |
|  | IU3^ | 4.25 | 0.73 |
| Social media use | SM1 | 4.26 | 0.73 |
|  | SM2^ | 4.48 | 0.77 |
|  | SM3 | 4.44 | 0.76 |
| Mobile device | MD1^ | 4.70 | 0.65 |
|  | MD2 | 4.22 | 0.98 |
|  | MD3 | 4.29 | 0.79 |
| Incident reporting | IR1^ | 4.23 | 0.69 |
|  | IR2^ | 4.25 | 0.64 |
|  | IR3 | 3.92 | 1.11 |

The study revealed varying levels of cybersecurity awareness across different domains. *Password management* showed high agreement, with PM3 scoring the highest. *Email usage* displayed diverse patterns, from EU1's low 3.70 to EU2's highest positive response. Internet usage scored positively (3.91–4.33), with IU2 showing high consistency (SD=0.63). Social media and mobile device usage both indicated positive behaviors with high scores. Incident reporting received moderately high scores overall, suggesting a positive attitude, though IR3 scored slightly lower at 3.92 with higher variability (SD=1.11). These findings highlight areas of strength and potential improvement in students' cybersecurity practices.

### 4.3 Focus Area descriptive statistics per study level and age group

In this section, the descriptive statistics for each focus area are presented in Table 3, categorized as either undergraduate (UG) or postgraduate (PG).

**Table 3.** Descriptive statistics for UG and PG groups

| Focus area | Question label (^ indicates a negatively worded question) | $N_{UG}$ = 158 | | $N_{PG}$ = 108 | |
|---|---|---|---|---|---|
|  |  | Mean | Std Dev | Mean | Std Dev |
| Password management | PM1^ | 3.93 | 1.04 | 4.28 | 1 |
|  | PM2 | 4.03 | 0.97 | 4.14 | 1.21 |
|  | PM3^ | 4.12 | 0.9 | 4.48 | 0.7 |
| Email usage | EU1^ | 3.46 | 1.1 | 4.06 | 0.94 |
|  | EU2^ | 4.85 | 0.37 | 4.81 | 0.46 |
|  | EU3 | 4.46 | 1.01 | 4.45 | 1.04 |
| Internet usage | IU1 | 3.91 | 0.91 | 3.87 | 0.94 |
|  | IU2 | 4.39 | 0.6 | 4.23 | 0.68 |
|  | IU3^ | 4.25 | 0.76 | 4.25 | 0.69 |
| Social media use | SM1 | 4.21 | 0.73 | 4.33 | 0.74 |
|  | SM2^ | 4.44 | 0.84 | 4.55 | 0.65 |
|  | SM3 | 4.44 | 0.75 | 4.5 | 0.77 |
| Mobile device | MD1^ | 4.68 | 0.62 | 4.73 | 0.69 |
|  | MD2 | 4.13 | 1.01 | 4.36 | 0.91 |
|  | MD3 | 4.3 | 0.75 | 4.26 | 0.86 |
| Incident reporting | IR1^ | 4.18 | 0.7 | 4.31 | 0.68 |
|  | IR2^ | 4.25 | 0.66 | 4.26 | 0.63 |
|  | IR3 | 3.99 | 1.01 | 3.82 | 1.23 |

The study revealed varying cybersecurity practices between UG and PG students across different domains. PG students demonstrated better *password management* attitudes (mean scores 4.14–4.48) compared to UG students (3.93–4.12). Similarly, PG students showed more positive *email usage* (EU) attitudes than UG students with the aggregate PG EU score showing ~4% better attitudes compared to UG (although the individual EU items show UG attitudes slightly exceed PG attitudes in two EU items). *Internet usage* attitudes were positive for both groups, with UG students scoring slightly higher. Both groups exhibited high scores in *social media* and *mobile device usage*, with PG students marginally outperforming UG students. Notably, despite overall high mean scores, both groups showed a significant reduction in the IR3 measure, indicating a common area for improvement regardless of academic level.

Analysis of cybersecurity practices across age groups and educational levels revealed varied patterns. In password management, both UG and PG students showed positive attitudes, with older PG students (30-50) demonstrating the strongest practices. Email usage varied, with UG students showing inconsistent behavior across age groups, while PG students, particularly those aged 30-40, exhibited more positive attitudes. Internet usage was generally positive across all age groups and educational levels. Social media usage was high among most UG age groups, with a slight dip in the 25-30 category, possibly due to sample size limitations. PG students consistently showed high social media awareness across all ages. Mobile device usage was high for both UG and PG students, with PG students scoring slightly higher, especially in MD1. Incident reporting was generally strong across UG age groups, while PG students showed high awareness in IR1 and IR2, but notable differences in IR3 for ages 25-30 and 35-40. These findings highlight age-related and educational study level differences in cybersecurity practices, suggesting areas for targeted improvements in awareness and education.

### 4.4 Validity and Reliability

Table 5 shows the reliability scores for the focus areas. The reliability testing for each focus area was validated [7]. The research instrument is underpinned by the KAB (Knowledge, Attitude, and Behavior) model and only the attitude construct with the focus area variables was assessed within this research using a modular approach tailored towards the focus areas in cybersecurity.

**Table 4.** Reliability scores for the focus areas area question

| S. No | Focus Area | Cronbach's alpha |
|---|---|---|
| 1. | Password Management | 0.82 |
| 2. | Email use | 0.78 |
| 3. | Internet use | 0.78 |
| 4. | Social Media use | 0.75 |
| 5. | Mobile Devices | 0.81 |
| 6. | Incident Reporting | 0.79 |

### 4.5 T-Test to compare study level groups

A T-test analysis comparing cybersecurity attitudes between UG ($N_{UG}$=158), and PG ($N_{PG}$=108) students revealed notable differences. PG students demonstrated more positive attitudes in 55.56% of variables, particularly in PM1^, PM3^, and EU1^, indicating better cybersecurity practices. UG students scored higher in 16.67% of variables, notably in IU2. No significant difference was found in 27.78% of variables. Statistically significant differences ($p < 0.05$) were observed in PM1^, PM3^, EU1^, and IU2, rejecting the null hypothesis for these variables. This analysis highlights varying cybersecurity awareness levels between UG and PG students, suggesting areas for targeted educational interventions to enhance overall cybersecurity practices among the student population.

## 5 Discussion and Recommendations

The primary objective of this research was to evaluate attitudes toward cybersecurity predispositions among undergraduate (UG) and postgraduate (PG) students, considering their educational status and age. Our findings reveal a significant *Cybersecurity-Resilience Gap* between UG and PG students, with postgraduates demonstrating more advanced awareness and practices across key dimensions of digital security.

Our analysis indicates that postgraduate students statistically outperformed undergraduates in overall cybersecurity awareness attitudes, particularly in password management and email use. This disparity suggests the existence of a *Cybersecurity-Resilience Inflection Point* that occurs during the transition from undergraduate to postgraduate studies, where students' cybersecurity awareness and practices markedly improve. Furthermore, we observed a positive correlation between age and improved cybersecurity attitudes, indicating a *Cybersecurity-Resilience Education* effect that develops over time and academic progression.

These insights underscore the critical need for enhanced Cybersecurity-Resilience Education in higher education institutions. To address the identified gaps, we recommend implementing comprehensive cybersecurity awareness programs for undergraduate students, with a particular focus on password management and email security. This early intervention can help narrow the Cybersecurity-Resilience Gap earlier in students' academic careers. Additionally, developing age-appropriate and education-level-specific cybersecurity training modules can accelerate the Cybersecurity-Resilience Inflection Point, enhancing students' resilience against cyber threats.

To reinforce the *Cybersecurity-Resilience Education* effect, we propose integrating ongoing cybersecurity education throughout academic programs. This approach should include practical exercises and real-world simulations to improve attitudes and behaviors. Extending this education to faculty and staff is crucial for ensuring a comprehensive approach to cybersecurity awareness across the institution.

To gain a broader understanding of the *Cybersecurity-Resilience Gap* and develop standardized educational interventions, we suggest expanding this study model to other South African universities. This cross-institutional collaboration can provide valuable insights into regional trends and best practices in cybersecurity education.

Addressing the *Cybersecurity-Resilience Gap* through targeted and continuous *Cybersecurity-Resilience Education* is crucial for developing a cyber-resilient academic community. By understanding and leveraging the *Cybersecurity-Resilience Inflection* Point, institutions can more effectively prepare students for the ever-changing nature of cyber threats. As our digital world continues to expand, fostering a culture of cybersecurity awareness becomes not just a necessity, but a fundamental aspect of modern education.

## 6      Conclusion

This study proposes a novel *Cybersecurity-Resilience Gap* between undergraduate and postgraduate students, with the latter demonstrating superior awareness and practices. Postgraduates' stronger cybersecurity attitudes and skills suggest a crucial inflection point during the transition from undergraduate to postgraduate studies. To bridge this gap, we recommend, firstly, an enhanced cybersecurity education to focus on practical training for undergraduates, particularly in password management and email security. Secondly, continuous education to integrate cybersecurity into academic programs across all levels. Lastly, an institutional-wide approach to extend training to faculty and staff for a comprehensive approach. Future research could expand this study to other universities, address limitations, and explore broader perspectives, to inform effective cybersecurity education strategies.